\documentclass[journal=jacsat,manuscript=special issue,14pt]{achemso}
\usepackage{amsfonts}
\usepackage{amsmath}
\usepackage{amssymb}
\usepackage{amsthm}
\usepackage{appendix}
\usepackage{datetime}
\usepackage{dcolumn} % Align table columns on decimal point
\usepackage{enumerate}
\usepackage[T1]{fontenc}
\usepackage{graphicx} % Include figure files
\usepackage[utf8]{inputenc}
\usepackage[unicode]{hyperref}
\hypersetup{
    colorlinks=true,
    linkcolor=blue,
    filecolor=magenta,
    urlcolor=cyan,
}
\usepackage{mathptmx}
\usepackage{tensor}
\usepackage{soul} % strike out with \st
\usepackage{xcolor}
\usepackage{yfonts}

\newcommand{\bfr}{{\bf r}}

\newcommand{\bfR}{{\bf R}}

\newcommand{\bra}[1]{\ensuremath{\langle #1 \vert}}
\newcommand{\ket}[1]{\ensuremath{\vert #1  \rangle}}
\newcommand{\expect}[2]{\ensuremath{\langle #2 \vert #1 \vert #2 \rangle}}
\newcommand{\op}[1]{\mathsf #1}

\DeclareMathOperator{\dd}{d}

\DeclareMathOperator{\erfc}{erfc}
\makeatletter
\newcommand{\thickbar}{\mathpalette\@thickbar}
\newcommand{\@thickbar}[2]{{#1\mkern1.5mu\vbox{
  \sbox\z@{$#1\mkern-1.5mu#2\mkern-1.5mu$}%
  \sbox\tw@{$#1\overline{#2}$}%
  \dimen@=\dimexpr\ht\tw@-\ht\z@-.8\p@\relax
  \hrule\@height.8\p@ % adjust for the desired rule thickness
  \vskip\dimen@
  \box\z@}\mkern1.5mu}
}
\makeatother
\renewcommand*{\bar}{\thickbar}
\newcommand{\anthony}[1]{{#1}}
\renewcommand{\st}[1]{{}}

\usepackage{achemso} % for \achemso
\SectionNumbersOn    % for referencing section, tableofcontent in achemso

\title{Long-range configuration interaction with an {\em ab initio} short-range correction and an asymptotic lower bound}

\author{Anthony Scemama}
\affiliation{Laboratoire de Chimie et Physique Quantiques (UMR 5626),
Université de Toulouse, CNRS, UPS, France}
\email{scemama@irsamc.ups-tlse.fr}

\author{Andreas Savin}
\affiliation{Laboratoire de Chimie Th\'eorique, CNRS and Sorbonne University \\ 4 place Jussieu, 75252 Paris, France}
\email{andreas.savin@lct.jussieu.fr}


\begin{document}

%\begin{flushright}
%\begin{minipage}{0.5\textwidth}
%\noindent
%\begin{small}
% submitted for publication in: \textit{J. Phys. Chem.}
%\end{small}
%\end{minipage}
%\end{flushright}

\begin{abstract}
Short-range corrections to long-range selected configuration interaction calculations are derived from perturbation theory considerations and applied to harmonium (with two to six electrons for some low-lying states).
No fitting to reference data is used, and the method is applicable to ground and excited states.
The formulas derived are rigorous when the physical interaction is approached.
In this regime, the second-order expression provides a lower bound to the \anthony{long-range full configuration interaction} energy.
A long-range/short-range separation of the interaction between electrons at a distance of the order of one atomic unit provides total energies within chemical accuracy, and, for the systems studied, provide better results than short-range density functional approximations.
\end{abstract}

\maketitle

\newpage
\tableofcontents
\newpage

\centerline{\today \hspace{2pt} at \currenttime}

\section{Introduction}

In this paper we perform selected configuration interaction (CI) calculations for Hamiltonians with long-range interaction between electrons, and correct for the missing term exploiting the universal behavior of the wave function at short range.
This is motivated by the reduction of computational effort: shorter CI expansions are needed when the interaction is weak.
The method presented here is related to previous work where the short range interaction was described by short-range density functionals (see, e.g., ~\citenum{StoSav-INC-85, SavFla-IJQC-95, Sav-INC-96, PolSavLeiSto-JCP-02, TouColSav-JCP-05}).
In a more recent development, we used an approach \anthony{\st{reminding} reminiscent} of the one used in generating (and motivating) density functional approximations, namely the adiabatic connection.
However, we do not construct any density functional, but only switch on the short-range interaction (see, {\em e.g.}, Refs.~\citenum{Sav-JCP-20, SceSav-JPC-24}).
This can be done for ground and excited states, and does not require a self-consistent improvement of the wave function.
There is no transfer of information from other systems (like the uniform electron gas, as done in density functional approximations) and there are no empirical parameters.
Instead, the exact, universal behavior of the wave function at very short inter-electronic distances is used in the regime when the physical interaction is approached.
The additional computational effort for the short-range correction is negligible in comparison to that for the CI calculation with the long-range interaction operator.

In this paper, we continue in this vein.
Here, we use a perturbational treatment to analyze the terms showing up.
Truncating at first order, we have the expectation value of the wave function.
This provides an upper bound.
At second order, we obtain a lower bound to the energy.
The examples show that it is superior to the upper bound.
At third order, we recover the adiabatic connection result.
In all examples studied, the errors are within chemical accuracy and better than those obtained using short-range density functionals.

\section{Method}

\subsection{Perturbation approach}
Instead of considering the Schr\"odinger equation for the electronic Hamiltonian, $\op{H}$, we consider an electronic Hamiltonian, $\op{H}(\mu)$, and \anthony{\st{correct} modify} it by eliminating the short range interaction from it:
\begin{equation}
    \label{eq:h-model}
    \op{H}(\mu) = \op{H} - \bar{\op{W}}(\mu).
\end{equation}
where
\begin{equation}
\label{eq:Wbar}
\bar{\op{W}}(\mu)  = \sum_{i=1}^{N-1} \sum_{j=i}^{N} \bar{w}(r_{ij}, \mu)
\end{equation}
and we choose
\begin{equation}
\label{eq:wbar}
\bar{w}(r,\mu)  = \frac{\erfc(\mu r)}{r}.
\end{equation}
In these equations, $r_{ij}$ represents the distance between electrons $i$ and $j$, and $\mu$ is a parameter that characterizes the range of the interaction.
It has the dimensions of an inverse length; the range of $\bar{w}$  becomes smaller as $\mu$ increases.
For $\mu=\infty$, the correction vanishes: $\op{H}(\mu=\infty)=\op{H}$.
We use a selected CI (very well approximating a full CI, FCI) calculation to estimate accurately the eigenvalue of $\op{H}(\mu)$, $E(\mu)$ and its normalized eigenfunction, $\Psi(\mu)$.
For $\mu=\infty$, we obtain the eigenvalue, $E=E(\infty)$, and eigenfunction, $\Psi=\Psi(\infty)$ of $\op{H}$.

Let us consider
\begin{equation}
    \label{eq:pert-H}
    \op{H}(\lambda,\mu) = \op{H}(\mu) + \lambda \bar{\op{W}}(\mu) ,
\end{equation}
that is, we consider $\bar{\op{W}}(\mu)$ as a perturbation.
We can obtain the energy from the adiabatic connection,
\begin{equation}
    \label{eq:ac}
    E = E(\mu)+ \int_0^1 \dd \lambda \, \partial_\lambda E(\lambda,\mu) = E(\mu)+ \int_0^1 \dd \lambda \, \expect{\bar{\op{W}}(\mu)}{\Psi(\lambda,\mu)} ,
\end{equation}
where $E(\lambda,\mu)$ is the eigenvalue, and $\Psi(\lambda,\mu)$ is the eigenfunction of $\op{H}(\lambda,\mu)$.
At large $\mu$, it becomes exact at first order in $\lambda$ (see appendix in Ref.~\citenum{SavKar-JCP-23}),
\begin{equation}
    \label{eq:pert-Psi}
    \Psi(\lambda,\mu) = \Psi(\mu) + \lambda \Psi^{(1)}(\mu) + \dots , \;\; \text{for large} \, \mu .
\end{equation}
The higher orders in $\lambda$ do not show up if we consider only the terms up to order $1/\mu$.
Note that $\Psi(0,\mu) = \Psi(\mu)$, and $\Psi(1,\mu)=\Psi$.

Instead of performing a standard perturbation theory,
let us use $\Psi(\lambda,\mu)$, eq.~\eqref{eq:pert-Psi} into eq.~\eqref{eq:ac}.
and retain the terms of the $|\Psi(\lambda,\mu)|^2$ up to a given power of $\lambda$, $k$.
We get the approximations to order $n$
\begin{equation}
    \label{eq:eok}
    E \approx E(\mu) + \bar{E}^{(k)}
\end{equation}
where
\begin{align}
    \label{eq:o1}
    \bar{E}^{(1)} & = \expect{\bar{\op{W}}(\mu)}{\Psi(\mu)} \\
    \label{eq:o2}
    \bar{E}^{(2)} & = \bar{E}^{(1)} + \bra{\Psi(\mu)} \bar{\op{W}}(\mu)\ket{\Psi^{(1)}(\mu)} \\
    \label{eq:o3}
    \bar{E}^{(3)} & = \bar{E}^{(2)} + \frac{1}{3} \expect{\bar{\op{W}}(\mu)}{\Psi^{(1)}(\mu)}
\end{align}
Using eq.~\eqref{eq:pert-Psi} at $\lambda=1$, $\Psi \approx \Psi(\mu)+\Psi^{(1)}(\mu)$, we get the following approximations, for order $k=0$ to $3$.
\begin{align}
    \label{eq:e0}
    E & \approx E(\mu) \\
    \label{eq:e1}
    E & \approx E(\mu) +  \expect{\bar{\op{W}}(\mu)}{\Psi(\mu)} \\
    \label{eq:e2}
    E & \approx E(\mu) + \bra{\Psi(\mu)} \bar{\op{W}}(\mu)\ket{\Psi}\\
    \label{eq:e3}
    E & \approx E(\mu) + \frac{1}{3} \left(
        \expect{\bar{\op{W}}(\mu)}{\Psi(\mu)}
        + \bra{\Psi(\mu)} \bar{\op{W}}(\mu)\ket{\Psi}
        + \expect{\bar{\op{W}}(\mu)}{\Psi}
     \right)
\end{align}
The adiabatic connection using $\Psi(\lambda,\mu)$ given by eq.~\eqref{eq:pert-Psi} yields eq.~\eqref{eq:e3}, because $\Psi(\lambda,\mu)$ is linear in $\lambda^2$.
Furthermore, we note that the correction to $\bar{E}^{(2)}$ in eq.~\eqref{eq:o2} is positive.
Thus, staying with the second order gives a lower bound to the result that would be obtained by adiabatic connection (see also eq.~(48) of Ref.~\citenum{Sav-JCP-20}).
We remind that these results are valid only for large $\mu$, as we have neglected the terms in $1/\mu^2$.
Eq.~\eqref{eq:e1} provides an upper bound to the energy, because it is equal to $\expect{\op{H}}{\Psi(\mu)}$.

Note the presence of $\Psi$ in eqs.~\eqref{eq:e2} and \eqref{eq:e3}.
The formulas seem useless.
However, as $\Psi$ appears only in the integrals containing a short-range operator, we can exploit the universal short-range behavior of $\Psi$ to calculate these integrals.

\subsection{Exploiting the short-range behavior of the wave function}

The Kato cusp condition~\cite{Kat-CPAM-57, KutMor-JCP-92, KurNakNak-AQC-16} gives $\Psi$ for  $r=|\bfr_{12}| \rightarrow 0$
\begin{equation}
\label{eq:psi-sr}
 \Psi(\bfR, \bfr) = \sum_{\ell,m} c_{\ell,m}(\bfR) \phi_{\ell}(r) Y_{\ell,m}(\Omega)
\end{equation}
where,
\begin{equation}
    \label{eq:Kato}
    \phi_{\ell}(r) = r^\ell \left[ 1 +\frac{1}{2 \ell+2 } r + \dots \right]
\end{equation}
$Y_{\ell,m}$ are spherical harmonics, $\Omega$ the solid angle associated to $\bfr_{12}$, and $\bfR$ denotes all the coordinates except $\bfr_{12}$.
For small $r$, to first order in $1/\mu$, one can find similar expressions~\cite{GorSav-PRA-06} also for $\Psi(\mu)$, the only change being that $\phi_\ell(r)$ is to replaced by $\phi_\ell (r, \mu)$, given explictly,  for example, in eq. (19) of Ref.~\citenum{SavKar-JCP-23}.

The $r^\ell$ factor in $\phi_\ell$ makes the wave function vanish for small $r$.
Therefore, in Ref.~\cite{SceSav-JPC-24}, we explored ignoring in eq.~\eqref{eq:psi-sr} the terms with $\ell>1$.
As a result, the pair functions, needed to obtain the expectation values showing up in the corrections to $E(\mu)$ are reduced to two components, one for $\ell=0$, one for $\ell=1$.
One obtains~\cite{SceSav-JPC-24}:
\begin{equation}
    \label{eq:W-generic}
    \bra{\Psi(\mu_1)} \bar{\op{W}}(\mu) \ket{\Psi(\mu_2)} = \mathcal{N}_s^2 \int_0^\infty \dd r \; r^2 \; \phi_0(r,\mu_1) \bar{w}(r) \phi_0(r,\mu_2) +
    \mathcal{N}_t^2 \int_1^\infty \dd r \; r^2 \; \phi_1(r,\mu_1) \bar{w}(r) \phi_1(r,\mu_2) .
\end{equation}
$\mathcal{N}_s^2$, and $\mathcal{N}_s^2$ appear through integration over $\bfR$ and $\Omega$.
$\mathcal{N}_{s,t}^2$ are determined by the exact wave function.~\cite{GorSav-PRA-06, Sav-JCP-20}
Thus, they don't depend on $\mu$, and can be re-constructed from information obtained for the model system~\cite{Sav-JCP-20},
\begin{align}
\label{eq:ws}
 \mathcal{N}_s^2 & = \frac{\expect{\bar{\op{W}}(\mu)}{\Psi(\mu)}_s}{\int_0^\infty \dd r \,  r^2 \, \vert \phi_0(r,\mu) \vert^2 \, \bar{w}(r,\mu)} \\
\label{eq:wt}
 \mathcal{N}_t^2 & = \frac{\expect{\bar{\op{W}}(\mu)}{\Psi(\mu)}_t}{\int_0^\infty \dd r \,  r^2 \, \vert \phi_1(r,\mu) \vert^2 \, \bar{w}(r,\mu)}.
\end{align}
The indices $s,t$ indicate the singlet or triplet components (Eqs.~(22) and (23) in Ref.~\citenum{SceSav-JPC-24}).

The integrals over $r$ appearing in eq.~\eqref{eq:W-generic} have the form:
\begin{multline}
    \label{eq:w-all}
    \int_0^\infty \dd r \;  r^2 \; \phi_\ell(r,\mu_1) \phi_\ell(r,\mu_2)  \;  \bar{w}(r,\mu)  = \\
        c_{2l+2}(\mu_1,\mu_2) \, \mu^{-(2l+2)} + c_{2l+3}(\mu_1,\mu_2) \, \mu^{-(2l+3)} + c_{2l+4}(\mu_1,\mu_2) \, \mu^{-(2l+4)}
\end{multline}
The values of $\mu_1$ and $\mu_2$ needed for $\bar{E}^{(k)}, \; k=1,2,3$ are $\mu$ and $\infty$ and $\ell=0, 1$.
The coefficients $c$ are given in tab.~\ref{tab:coefwb}.
%\begin{align}
%   \int_0^\infty \dd r \; r^2 \; \vert \phi_0(r,\mu) \vert^2 \; \bar{w}(r,\mu) & =  0.25/\mu^{2}  + 0.343860/\mu^3  + 0.121951/\mu^4 \\
%    \int_0^\infty \dd r \;  r^2 \;  \vert \phi_1(r,\mu) \vert^2 \;  \bar{w}(r,\mu)  & =   0.1875/\mu^4  + 0.126527/\mu^5  +  0.022997/\mu^6 \\
%   \int_0^\infty \dd r \;  r^2 \;  \phi_0(r) \phi_0(r,\mu)  \;  \bar{w}(r,\mu) & =   0.25/\mu^2 + 0.265962/\mu^3 + 0.071039/\mu^4 \\
%    \int_0^\infty \dd r \;  \phi_1(r) \phi_1(r,\mu)  \;  \bar{w}(r,\mu)  & = 0.1875/\mu^4 + 0.119683/\mu^5 + 0.021075/\mu^6  \\
%    \int_0^\infty \dd r \; r^2 \;  \vert \phi_0(r) \vert^2 \;  \bar{w}(r,\mu)  & = 0.25/\mu^2 +0.188063/\mu^3 + 0.046875/\mu^4 \\
%    \int_0^\infty \dd r \;  r^2 \;  \vert \phi_1(r) \vert^2 \;  \bar{w}(r,\mu)  & = 0.1875/\mu^4 + 0.112838/\mu^5 + 0.019531/\mu^6  .
%\end{align}
\begin{table}[]
    \centering
%    \begin{ruledtabular}
    \begin{tabular}{llllll}
\hline
\hline
    $\mu_1$ & $\mu_2$ & $\ell$ & $c_{2l+2}$ & $c_{2l+3}$  & $c_{2l+4}$ \\ \hline
    $\mu$    & $\mu$    & 0 & 0.25   & 0.343860   &  0.121951  \\ \hline
    $\mu$    & $\mu$    & 1 & 0.1875 & 0.126527   &  0.022997  \\ \hline
    $\mu$    & $\infty$ & 0 & 0.25   & 0.265962   &  0.071039  \\ \hline
    $\mu$    & $\infty$ & 1 & 0.1875 & 0.119683   &  0.021075  \\ \hline
    $\infty$ & $\infty$ & 0 & 0.25   & 0.188063   &  0.046875  \\ \hline
    $\infty$ & $\infty$ & 1 & 0.1875 & 0.112838   &  0.019531  \\
\hline
\hline
    \end{tabular}
%    \end{ruledtabular}
    \caption{Coefficients appearing in the integrals $\int_0^\infty \dd r \;  r^2 \;  \phi_0(r,\mu) \phi_0(r,\mu_2)  \;  \bar{w}(r,\mu)$, eq.~\eqref{eq:w-all}.}
    \label{tab:coefwb}
\end{table}

We can merge the terms containing the integrals of the type given in eq.~\eqref{eq:w-all}, obtaining simple
expressions for $\bar{E}^{(k)}, k=1,2, \text{ or } 3$:
\begin{equation}
    \label{eq:working}
    E \approx \expect{\op{H}(\mu)}{\Psi(\mu)} + \alpha_s^{(k)}(\mu) \expect{\bar{\op{W}}(\mu)}{\Psi(\mu)}_s + \alpha_t^{(k)}(\mu) \expect{\bar{\op{W}}(\mu)}{\Psi(\mu)}_t
\end{equation}
The expressions for $\alpha_{s,t}$ as functions of $\mu$ are given in Table~\ref{tab:alphas}.
\begin{table}[htb]
    \centering
%\begin{ruledtabular}
\begin{tabular}{lcc}
\hline
\hline
$k$ & $\alpha_s^{(k)}(\mu)$ & $\alpha_t^{(k)}(\mu)$
  \\ \hline
0 & 0 & 0
  \\ \hline
1 & 1 & 1
  \\ \hline
  \\[-1cc]
2 & $\displaystyle \frac{\displaystyle \mu ^2+1.06385 \mu +0.284155}{\displaystyle \mu ^2+1.37544 \mu +0.487806}$
  & $\displaystyle \frac{\displaystyle \mu ^2+0.638308 \mu +0.112402}{\displaystyle \mu ^2+0.674813 \mu +0.122652}$
  \\[-1cc]
  \\ \hline
  \\[-1cc]
$3$ & $\displaystyle \frac{\displaystyle \mu ^2+1.06385 \mu +0.319820}{\displaystyle \mu ^2+1.37544 \mu +0.487806}$
  & $\displaystyle\frac{\displaystyle \mu ^2+0.638308 \mu +0.113074}{\displaystyle \mu ^2+0.674813 \mu +0.122652}$ \\
  \\[-1cc]
\hline
\hline
\end{tabular}
%\end{ruledtabular}
    \caption{Functions $\alpha_{s,t}^{(k)}$ showing up in eq.~\eqref{eq:working}.}
    \label{tab:alphas}
\end{table}

For $\mu \rightarrow \infty$, $\alpha_{s,t}^{(k)} \rightarrow 1$, that is, $E(\mu \rightarrow \infty) \rightarrow \expect{\op{H}}{\Psi}=E$.
When a finite basis set is used, $E(\mu \rightarrow \infty)$ produces the energy of the system with physical interaction in this basis set.
We see from the expressions given in tab.~\ref{tab:alphas} that $\alpha_{s,t}^{(2)} < \alpha_{s,t}^{(3)}$ meaning that the lower bound property is satisfied.
Note that the $\alpha_{s,t}^{(k)}$ always remain finite, and monotonically increase with $\mu$.
This property was derived for large $\mu$.
For small $\mu$,  it can be \anthony{\st{shownthat} shown that} $\alpha_{s,t}$ should decrease with $\mu$.~\cite{SceSav-JPC-24}
As $\alpha_{s,t}^{(k)}(\mu=0) \in [0,1]$ there never is a divergence in $E^{(k)}$ that can be imputed to degeneracy in the zeroth order of perturbation theory.

At this point we would like to remind that with density functional approximations there is a discussion about using exchange and correlation together, or correlation only.
The argument uses not only the quality observed for the results with the different approximations.
The argument for using only correlation is based on the expectation that exchange is easily computed for molecules, and smaller errors will appear when approximating a smaller part of the energy.
An argument against it goes as follows.~\cite{LanPer-PRB-77}
Most density functional approximations use the properties of the uniform electron gas.
There one finds a compensation of long-range correlation by long-range exchange.
Thus, no spurious long-range correlation shows up in molecules when one also includes long-range exchange in the density functional.
It has been noted that when using short-range density functionals the difference between using exchange-correlation and correlation-only density functional approximations is not important.~\cite{SavFla-IJQC-95}
Because $\expect{\bar{\op{W}}(\mu)}{\Psi(\mu)}=\expect{\bar{\op{W}}(\mu)}{\Psi(\mu)}_s+\expect{\bar{\op{W}}(\mu)}{\Psi(\mu)}_t$, instead of correcting $E(\mu)$ one can correct $\expect{\op{H}}{\Psi(\mu)}$.
In the latter case, one has to replace $\alpha_{s,t}^{(k)}(\mu)$ by $\alpha_{s,t}^{(k)}(\mu)-1$.
Thus, in our approach, there is strictly no difference between correcting $E(\mu)$, or the expectation value of $\op{H}, E(\mu) + \bar{E}^{(1)}$.

Note that in our derivation we never assumed to treat the ground state.
The formulas can be used for excited states, too.

\subsection{The system studied}
In this paper we study $N=2, \dots, 6$ electrons confined by a harmonic potential, $v(\bfr)=(1/2) \omega^2 r^2$, where $\omega=1/2$.
Except for the ground state of the system with two electrons,\cite{Kin-TCA-96} no exact energies are known.
For the other systems, we compare with the best results we found in the literature, \anthony{obtained with explicitly correlated wave functions}.~\cite{CioStrMat-JCP-12,CioStrMat-JCP-14,CioStr-JCP-17,CioStr-JCP-18,Str-JCP-16}

We use the data from previous calculations presented in Ref.~\citenum{SceSav-JPC-24},
where our objective was to achieve the full configuration interaction (FCI) level of accuracy
utilizing \anthony{\st{an even-tempered Gaussian basis set} a 9s7p5d3f3g1h1i Gaussian basis set, given in Table~1 of
ref.}~\citenum{SceSav-JPC-24}.
For $N=2$, FCI calculations were performed for all $\mu$ values.
For larger systems, we employed the Configuration Interaction using a Perturbative
Selection made Iteratively (CIPSI) algorithm,\cite{HurMalRan-73} aiming to
attain the highest feasible approximation to FCI while maintaining a
minimized computational cost.
As the FCI wave function can be expressed using any set of orbitals, and as the
wave functions are close \anthony{to the \st{the}} FCI \anthony{wave functions},
the orbitals and the set of selected determinants was kept constant as those obtained with $\mu=\infty$, and
the wave function $\Psi(\mu)$ was obtained by diagonalizing $\op{H}(\mu)$ in the basis of selected determinants.
All the configuration interaction calculations were performed using the \textsc{Quantum Package} software.\cite{GarGasBen-JCTC-19}
The results presented in tables were obtained with Mathematica.\cite{Wol-23}

\anthony{In the three-dimensional harmonic oscillator the ordering of orbital
energy levels is $\varepsilon_{1s} < \varepsilon_{1p} <
\varepsilon_{2s}$ that is different from that in atoms where level
ordering is $\varepsilon_{1s} < \varepsilon_{2s} < \varepsilon_{2p}$.
Thus, for example for $N=6$, we describe a system that resembles more to
the valence shell of the oxygen atom, and not to that of the carbon atom.}

\section{Results}

\subsection{Energy errors}

\begin{figure}[htbp]
    \centering
    \includegraphics[width=0.5\textwidth]{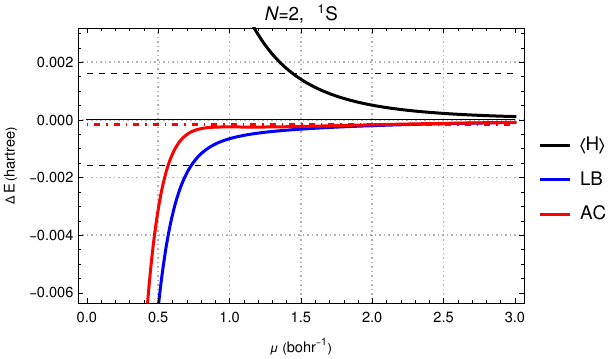} \\
    \includegraphics[width=0.5\textwidth]{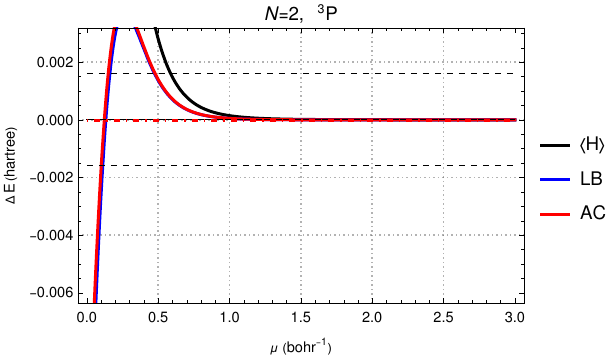} \\
    \includegraphics[width=0.5\textwidth]{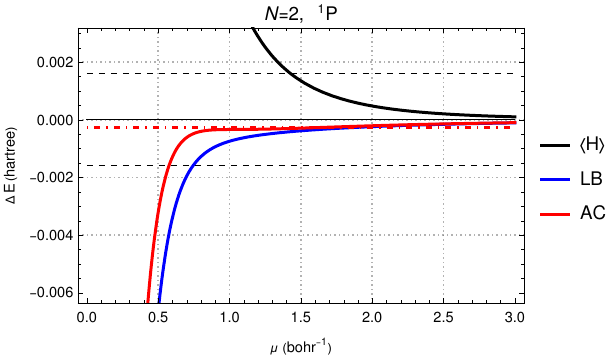}
    \caption{\anthony{\st{Energy errors} Differences between our approximations and our FCI energy} (in hartree) for harmonium with $N=2$ electrons, as a function of the model parameter $\mu$ (in bohr$^{-1}$) for the $^1$S ground state, $^3$P and $^1$P excited states. The model energies are corrected up to first order, ($\expect{\op{H}}{\Psi(\mu)}$, eq.~\eqref{eq:e1}, black curves), second order (asymptotic lower bound (LB)  eq.~\eqref{eq:e2}, blue curves), and third order (adiabatic connection (AC), eq.~\eqref{eq:e3},  red curves). The black horizontal dashed lines mark the domain within which the error with respect to the reference full configuration interaction (FCI) calculation is within chemical accuracy. The horizontal red dot-dashed lines show the difference between the best energy from the literature and the FCI energy.}
    \label{fig:de-n2}
\end{figure}

\begin{figure}[htbp]
    \centering
    \includegraphics[width=0.6\textwidth]{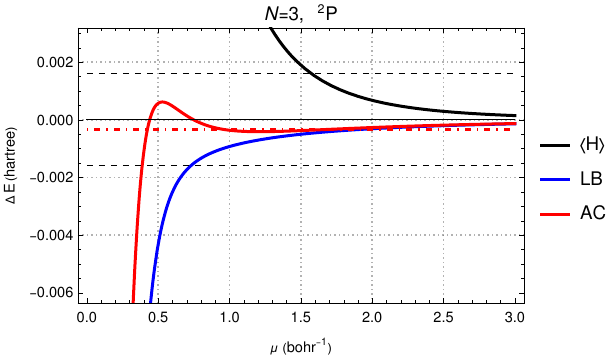} \\
    \includegraphics[width=0.6\textwidth]{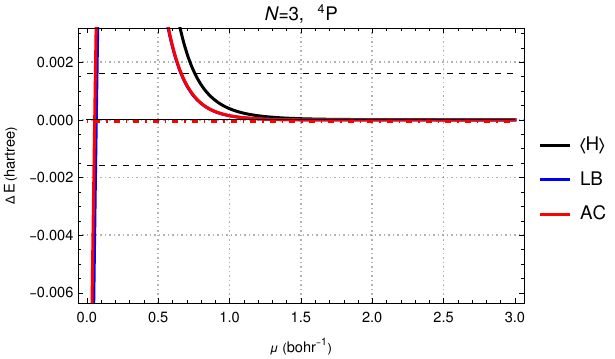} \\
    \includegraphics[width=0.6\textwidth]{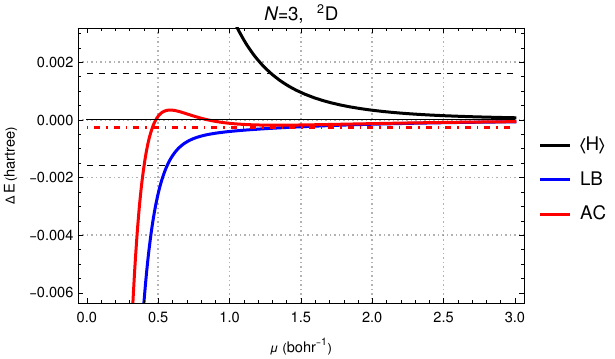}
    \caption{\anthony{\st{Energy errors} Differences between our approximations and our FCI energy} for harmonium with $N=3$ electrons for the $^2$P ground state, the $^4$P and $^2$D excited states. The color coding is the same as in fig.~\ref{fig:de-n2}.}
    \label{fig:de-n3}
\end{figure}

\begin{figure}[htbp]
    \centering
    \includegraphics[width=0.46\textwidth]{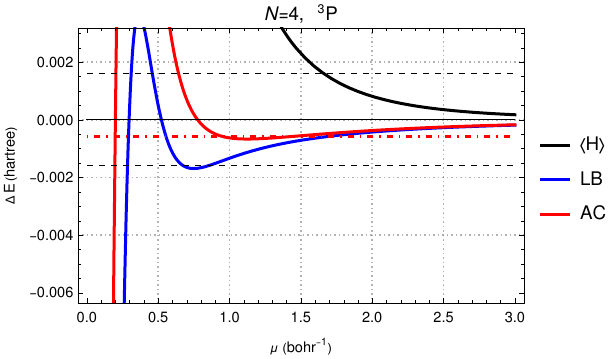}
    \includegraphics[width=0.46\textwidth]{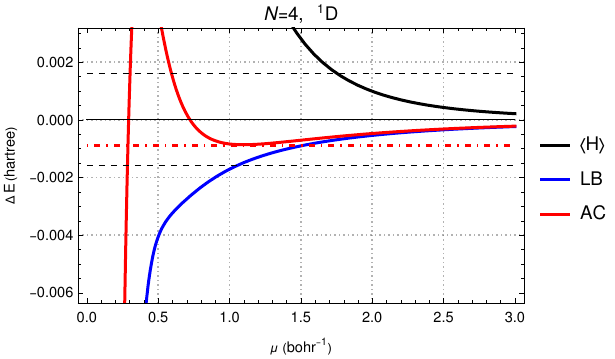} \\
    \includegraphics[width=0.46\textwidth]{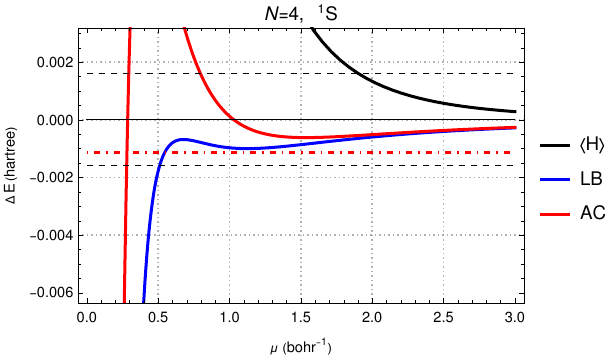}
    \includegraphics[width=0.46\textwidth]{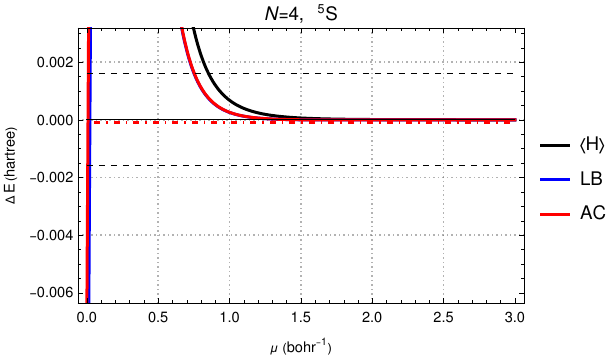}
    \caption{\anthony{\st{Energy errors} Differences between our approximations and our FCI energy} for harmonium with $N=4$ electrons for the $^3$P ground state, the $^1$D, $^1$S, and $^4$S excited states. The color coding is the same as in fig.~\ref{fig:de-n2}.}
    \label{fig:de-n4}
\end{figure}

\begin{figure}[htbp]
    \centering
    \includegraphics[width=0.6\textwidth]{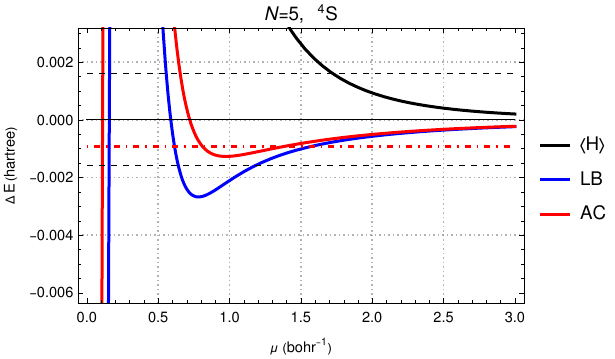} \\
    \includegraphics[width=0.6\textwidth]{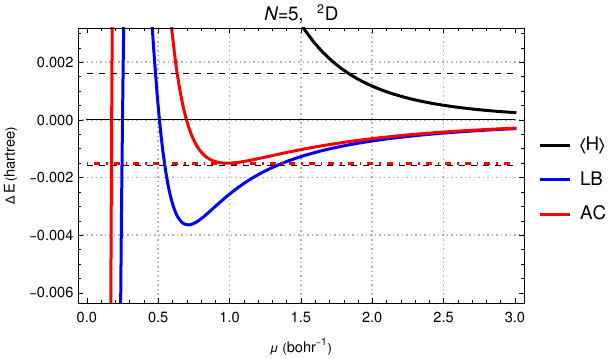} \\
    \includegraphics[width=0.6\textwidth]{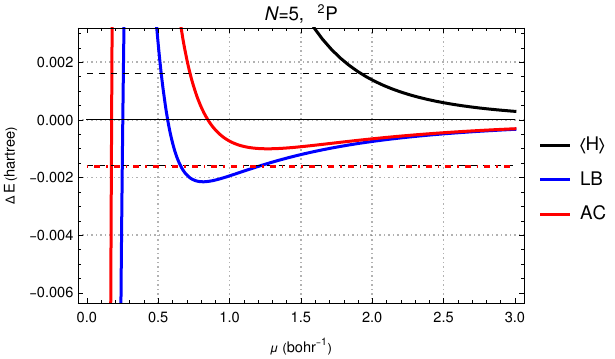}
    \caption{\anthony{\st{Energy errors} Differences between our approximations and our FCI energy} for harmonium with $N=5$ electrons for the $^5$S ground state, the $^2$D and $^2$P excited states. The color coding is the same as in fig.~\ref{fig:de-n2}.}
    \label{fig:de-n5}
\end{figure}

\begin{figure}[htbp]
    \centering
    \includegraphics[width=0.46\textwidth]{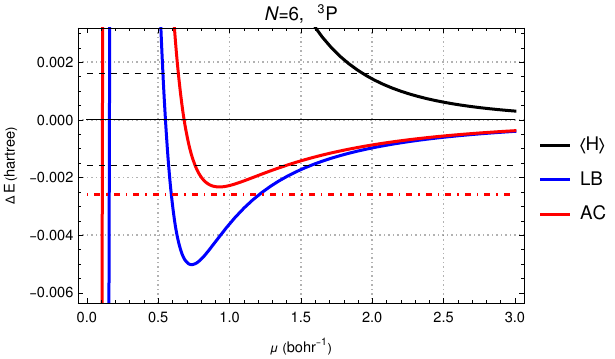}
    \includegraphics[width=0.46\textwidth]{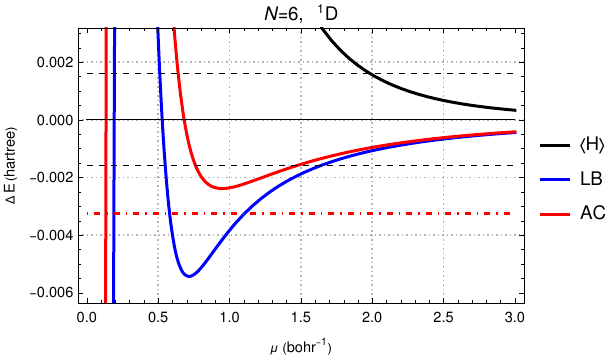}
    \caption{\anthony{\st{Energy errors}Differences between our approximations and our FCI energy} for harmonium with $N=6$ electrons for the $^3$P ground state, and the $^1$S excited state. The color coding is the same as in fig.~\ref{fig:de-n2}.}
    \label{fig:de-n6}
\end{figure}

The approach we presented above has two limitations:
i) it is valid only as the model interaction approaches the physical interaction (for large $\mu$),
ii) the basis set limitations are more severe when the interaction is stronger (the basis set errors are smaller when $\mu$ is small).
In order to explore the range of validity of our approximation we show in figs.~\ref{fig:de-n2} to \ref{fig:de-n6} the errors made by our approximations\anthony{\st{ as a function of $\mu$}.
The curves present $\Delta E$, the difference between the total energy computed at a given value of $\mu$ and that of the FCI calculation. Also, a dot-dashed line indicates the difference between the best estimate from literature, and our FCI calculation.}

We aim for errors within chemical accuracy~\cite{Pop-RMP-99}, that is within 1~kcal/mol of the best available energy.~\cite{Kin-TCA-96,CioStrMat-JCP-12,CioStrMat-JCP-14,CioStr-JCP-17,CioStr-JCP-18,Str-JCP-16}
We see that in all plots we get a good accuracy for surprisingly large $\mu$.
However, as we could expect, even our best approximation worsens dramatically for small $\mu$.
(In our examples, the change occurs for $\mu$ between 0.5 and 1.0~bohr$^{-1}$.)

The errors of the zeroth order approximation, $E(\mu)$ are large, and not shown.
They can be found in Ref.~\citenum{SceSav-JPC-24}.
Going to first order, $\expect{\op{H}}{\Psi(\mu)}$ one finds an improvement.
The second order correction, eq.~\eqref{eq:e2}, improves significantly the results.
With it we obtained a lower bound that (for large $\mu$) is closer to the exact result than the variational upper bound.
The results are further improved by making the third order correction, eq.~\eqref{eq:e3}.

At large $\mu$, the method approaches by construction the full configuration interaction (FCI) result.
For $N=2$, fig.~\ref{fig:de-n2}, or $N=3$, fig.~\ref{fig:de-n3}, where the FCI result is very close to the exact result, the errors stay close to zero, for large enough $\mu$.
%Below some value of $\mu < 1$, first the second-order correction, and then the third order correction worsens dramatically.

For $N>3$, the basis set error starts to be clearly seen.
As $\mu$ is reduced, the basis set quality is improving.
The best compromise between eliminating the basis set dependence and reducing $\mu$ is found -- in the systems studied -- around $\mu=1$~bohr$^{-1}$.
%When the basis set error gets too large, even for smaller $\mu$, the correction is not able to take full advantage.
We see that for some $\mu<0.5$~bohr$^{-1}$ the lower bound property of $E^{(2)}$ can be lost.
Sometimes, it may seem that, the second-order correction seems to work better, for example for $N=5$, in the $^2$P~state.
This could be related to the failure of the large $\mu$ expansion before reaching the point where there is no longer any basis set error.

\clearpage
\subsection{Error estimates}

\begin{figure}[htbp]
    \centering
    \includegraphics{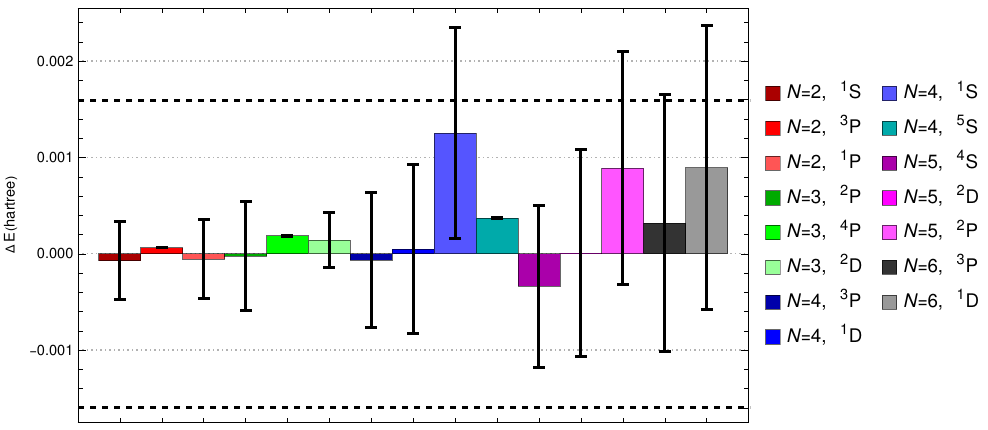}
    \caption{Errors, at $\mu=1$~bohr$^{-1}$, made by the third order correction (eq.~\eqref{eq:e3}, or adiabatic connection) with respect to the best estimates found in the literature~\cite{Kin-TCA-96,CioStrMat-JCP-12,CioStrMat-JCP-14,CioStr-JCP-17,CioStr-JCP-18,Str-JCP-16}; the error bars are given by the absolute value of the difference to the second order correction (eq.~\eqref{eq:e2}, or asymptotic lower bounds).}
    \label{fig:errorbars}
\end{figure}

Fig.~\ref{fig:errorbars} shows the errors after applying the third-order correction, that is, using the adiabatic connection in the regime approaching the Coulomb interaction.
\anthony{This time, we \st{We}} use as reference \anthony{(zero)} the best estimates available data in literature.~\cite{Kin-TCA-96,CioStrMat-JCP-12,CioStrMat-JCP-14,CioStr-JCP-17,CioStr-JCP-18,Str-JCP-16}
Although the formulas were derived to become exact for large values of the range-separation parameter, $\mu$, we present the results at $\mu=1$, in order to reduce the basis set errors.
In all cases studied,  the errors of the second-  and third-order correction (eqs.~\eqref{eq:e2} and \eqref{eq:e3}, respectively) are within chemical accuracy (using as reference the best values found in literature).
In addition to the values (already present in Ref.~\citenum{SceSav-JPC-24}) we show error estimates, based upon the difference between the second order estimate, that is the lower bound.
We also use this difference, with reversed sign, to define an upper bound.
\anthony{We notice that this error estimate is reasonable.}
The rigorous upper bound, the first-order estimate (the expectation value of the Hamiltonian) is not used in fig.~\ref{fig:errorbars} , because it gives often a too large error estimate.
The zero-order results (the uncorrected results) provide lower bounds.
However, the errors are huge in comparison with the estimate of the second-order correction, and not presented here.
We notice that in most cases, the reference value lies within the error bounds.
Notable exceptions to this rule are found for the states  having maximal spin multiplicity ($N=2, ^3$P; $N=3, ^4$P; $N=4, ^5$S), where the electrons are kept apart by the Pauli principle, and the corrections are small.
Nevertheless, the errors remain small in these cases.

\subsection{Comparison with density functional corrections}

\begin{figure}[htbp]
    \centering
    \includegraphics{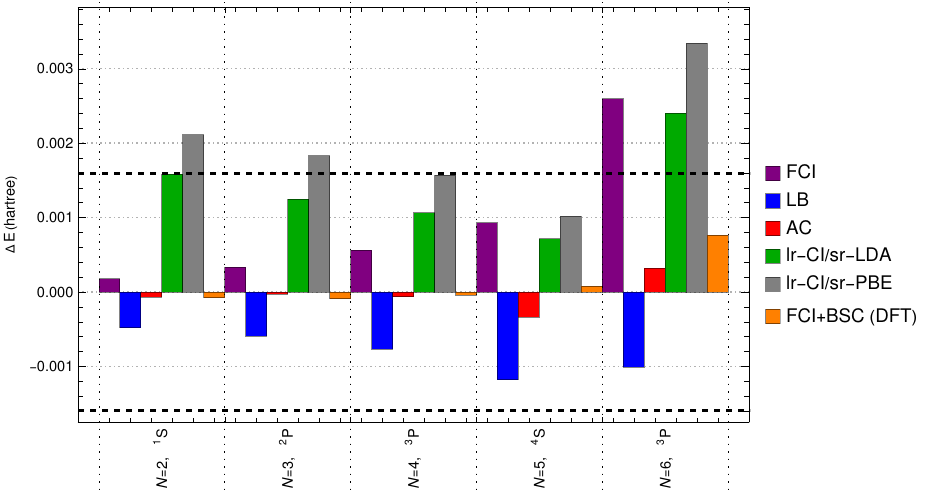} \\
    \caption{Errors, at $\mu=1$~bohr$^{-1}$, with respect to the best values found in the literature~\cite{Kin-TCA-96,CioStrMat-JCP-12,CioStrMat-JCP-14,CioStr-JCP-17,CioStr-JCP-18,Str-JCP-16} made by the full configuration calculation with Coulomb interaction (purple bars), by the second order correction (asymptotic lower bound, eq.~\eqref{eq:e2} bars), by the third order correction (adiabatic connection, eq.~\eqref{eq:e3}, red bars), by self-consistent calculations using a short-range density functional correction\cite{PerBurErn-PRL-96,Sav-INC-96,ZecGorMorBac-PRB-04, GolWerSto-PCCP-05,GolWerStoLeiGorSav-CP-06} (LDA, short-range local density approximation , green bars, or PBE, short range Perdew-Burke-Ernzerhof approximation, gray bars).
    Also shown are the results using density functional corrections of the basis set errors (orange bars).\cite{GinPraFerAssSavTou-JCP-18,LooPraSceTouGin-JPCL-19,GinSceLooTou-JCP-20,TraGinTou-JCP-23,SceSav-JPC-24}. The horizontal dotted lines delimit the region of chemical accuracy ($\pm~1$~kcal/mol).}
    \label{fig:barchart}
\end{figure}

One may wonder if similar results could be obtained using short-range density functionals.
In fig.~\ref{fig:barchart} we compare our results for ground states with those obtained in calculations combining long-range configuration interaction with short-range density functional approximations (lr-CI/sr-DFA) might be pertinent, as this is defined to cover all the domain of $\mu$.
For DFAs, we considered the local density approximation, LDA, and that of Perdew, Burke, and Ernzerhof, PBE~\cite{PerBurErn-PRL-96}, both adapted for short-range~\cite{Sav-INC-96,ZecGorMorBac-PRB-04, GolWerSto-PCCP-05,GolWerStoLeiGorSav-CP-06}.
We see that in every case our second- and third-order approximation outperforms the density functional approximations.
At first, this is surprising.
If the short-range behavior at first order is system-independent, as suggested by eq.~\eqref{eq:psi-sr}, why should a transfer from the uniform electron gas (as provided by the use of short-range LDA) give less accurate results?
Our guess is that the DFA attempts also to improve the results at small values of $\mu$, where our approximation is much worse.
Doing this transition, it worsens the quality of results in an uncontrolled way in the intermediate regime.
This also warns us that trying to improve the correction at smaller values of $\mu$ may hide some difficulties.
Another possible explanation is the treatment of open-shells (automatically present in the long-range configuration interaction calculation, that is not clearly defined in the density functional context.

We did not compare with density functional calculations for the excited states.
The errors of density functional calculations become larger.
This might be due to the use of ground-state density functional approximations.

We further see in fig.~\ref{fig:barchart} that the errors after the third-order correction are smaller than those of the reference full configuration interaction (due to the basis set) that are sometimes larger than 1~kcal/mol.
Using density functional basis set corrections~\cite{GinPraFerAssSavTou-JCP-18,LooPraSceTouGin-JPCL-19,GinSceLooTou-JCP-20,TraGinTou-JCP-23,SceSav-JPC-24} brings errors comparable to those with those of the third -order correction~\cite{SceSav-JPC-24}.
This can be rationalized by the use of larger values for the range-separation parameter ($\mu \approx 2-3$~bohr$^{-1}$): the density functional used contributes less, and is in a regime closer to the physical, Coulomb interaction.

\section{Summary and perspectives}

We start by solving accurately the Schr\"odinger equation where we have replaced the Coulomb interaction between electrons by a long-range interaction.
Short-range corrections are derived from perturbation theory, for the regime when the Coulomb interaction is approached.
The method is applied to some low-lying states of harmonium with two to six electrons.

\anthony{\st{For all systems studied, when $\mu$ decreases to about one atomic unit, the results improve, as basis set errors become less important.
We found in our examples that the results are not reliable for values of $\mu$ smaller than $0.5-1$~bohr$^{-1}$.
We do not have an explanation why an expansion in $1/\mu$, used as starting point for our approximation, still works around $\mu=1$.}
When the range separation parameter approaches infinity, we reach the
FCI result. However, for smaller values of $\mu$, the basis set error
(present in FCI) decreases. We notice that, for all systems studied,
when $\mu$ decreases to about one atomic unit, the results improve.
We found in our examples that the results degrade significantly for
$\mu$ between 0 and 0.5 -- 1 bohr$^{-1}$. One explanation of this effect can be
found in Ref.~\citenum{KarSav-PTC-24}.}

The third order correction, eq.~\eqref{eq:e3}, turns out to be identical to an approximation obtained previously using the adiabatic connection ($E_{L=1}$ in Ref.~\citenum{SceSav-JPC-24}).
It approaches the full configuration interaction, FCI, energy as the model system (with long-range interaction) gets closer to the physical system (with Coulomb interaction).
The second-order correction, eq.~\eqref{eq:e2}, provides a lower bound to the third-order energy.
Using a first order correction to the model energy, we obtain the expectation value of $\op{H}$, and thus an upper bound to the energy.
In all examples studied, we can see that for intermediate values of the range-separation parameter, $\mu$, we get better estimates of the third-order estimate by using the lower bound (second-order) estimate than by using the upper bound (first-order estimate).

The choice of $\mu=1$~bohr$^{-1}$ is reasonable, but arbitrary.
Furthermore, we may want to reduce the computational effort by reducing $\mu$.
One strategy would be to improve the description at small values of $\mu$.
For example, the local one-particle potential might be optimized.
(We have not done it in this paper, as we found that it is not relevant for large values of $\mu$ where the universal character of the wave function is dominant, and the modification of the external potential does not affect significantly the results in the region where the error is within chemical accuracy.~\cite{SceSav-JCC-24}
However, it is clear that it has an effect at small values of $\mu$.
Constructing a rational approximation satisfying some expressions for small $\mu$ and large $\mu$ might be helpful.

\anthony{Atoms and molecules can also be treated by the procedure described in
this paper. They are different due to the non-uniform character in these
systems (shells in atoms, different atoms in molecules). Should we keep
the simple, global treatment as used here for harmonium, or should
prefer a local treatment? The problematic is similar to that of the use
of plane waves (that extended over the whole system) versus that of
Gaussian basis sets (that are localized).}

A further limitation of our approach is that the optimal value of $\mu$ shows some system-dependence.
This can become critical if one wants to improve both the efficiency and the accuracy of the method.
We believe that a  {\em local} range-separation parameter might bring an improvement.
This could be done in the spirit of density-functional corrections~\cite{HenJanScuSav-IJQC-09,BruBahKum-JCP-22,BruBahKum-JCP-24} in long-range configuration interaction combined with short-range density functional approximations, or of basis set corrections~\cite{GinPraFerAssSavTou-JCP-18}.

\anthony{
The correction factors $\alpha_{s,t}^{(k)}$ showing up in eq.~\eqref{eq:working} and tab.~\ref{tab:alphas} are independent of the basis set used. However, the expectation values of $\op{H}$ and $\bar{\op{W}}$ in eq.~\eqref{eq:working} are basis set dependent.
In all our derivation, we have assumed that the problem is accurately solved for the model system (at $\mu$).
In practice, this is not the case.
In our numerical examples, even using a large basis set, we have seen that the FCI calculation may have errors larger than chemical accuracy.
Although the basis set error decreases at lower values of the range separation parameter $\mu$, the role of the basis set needs a separate, careful study.
}

\section*{Dedication}

This paper is dedicated to Gustavo Scuseria with whom it was always a pleasure to collaborate \dots and enjoy good food.

\section*{Acknowledgement}

We are grateful to Emmanuel Giner for the insightful discussions.
This work was realized using HPC resources from CALMIP (Toulouse) under allocation \texttt{p22001}.

\bibliography{biblio}
\end{document}